\documentclass{appolb}
\usepackage{graphicx}
\begin{document}
\title{Nuclear structure theory of the heaviest nuclei
\thanks{Presented at Zakopane Conference on Nuclear Physics ``Extremes
of the Nuclear Landscape''
}%
}
\author{A.\ V.\ Afanasjev and S.\ E.\ Agbemava}
%\address{Mississippi State University}
%\\
%{Third Author of different affiliation
%}
%the Name(s) of other Author(s)
%\address{affiliation}
%}
\maketitle
\begin{abstract}
  The current status of the application of covariant density functional 
theory to the description of actinides and superheavy nuclei is reviewed. 
The  achievements and open problems are discussed.
\end{abstract}

%%%%%%%%%%%%%%%%%%%%%%%%%%%%%%%%  
\section{Introduction}
%%%%%%%%%%%%%%%%%%%%%%%%%%%%%%%%

   There is a considerable activity (both in theory and experiment) in the 
study of shell-stabilized superheavy nuclei (SHN) \cite{SP.07,HG.08}. 
These studies are characterized by a number of experimental and theoretical 
challenges. The experimental challenges are the results of the experiments with low production 
cross sections and analyses based only on few events. The theoretical 
challenges are in part the consequences of different predictions for the centers
of the island of stability of SHN. Macroscopic+microscopic (MM) method, non-relativistic  
density functional theories (DFT) and covariant DFT (further CDFT 
\cite{VALR.05}) predict these centers at different proton and neutron numbers.
For example, these islands are predominantly centered at $(Z=114,N=184)$ and 
$(Z=126,N=184)$ in the MM and Skyrme DFT, respectively \cite{SP.07,A250}.
On the contrary,  covariant energy density functionals (CEDF's) predict
large shell gap at $Z=120$; however, neutron gap can be localized either
at $N=172$ (in most of the cases) and/or at $N=184$ \cite{BRRMG.99,A250,LG.14}.
In this situation it is important to understand the sources of the differences
and uncertainties in the prediction of the shell structure of SHN and how they
affect the physical observables (deformations, fission and $\alpha$-decay
observables) of interest. The actinides (the heaviest nuclei for which detailed spectroscopic
and fission information exists) play here a role of testing ground for the
state-of-the-art nuclear structure models.  We focus here on the results 
obtained with CDFT during last five years.

%%%%%%%%%%%%%%%%%%%%%%%%%%%%%%%%%%%%%
\section{Single-particle structure}
%%%%%%%%%%%%%%%%%%%%%%%%%%%%%%%%%%%%%

\begin{figure}[ht]
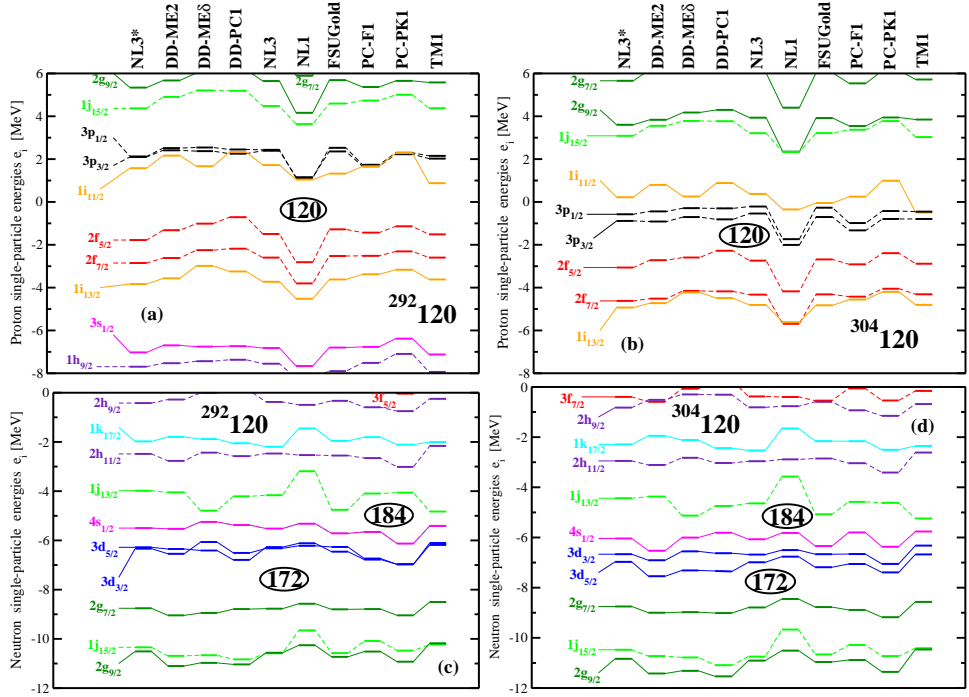

\includegraphics[angle=0,width=6.2cm]{fig-1-a.eps}
\includegraphics[angle=0,width=6.2cm]{fig-1-b.eps}
\includegraphics[angle=0,width=6.2cm]{fig-1-c.eps}
\includegraphics[angle=0,width=6.3cm]{fig-1-d.eps}
\caption{Neutron (left panels) and proton (right panels) 
single-particle states at spherical shape in the 
$^{292}$120 and $^{304}$120 SHN. They are determined with the 
indicated CEDF's in the RMF calculations without pairing. 
Solid and dashed connecting lines are used for positive and 
negative parity states. Spherical gaps are indicated; all 
the states below these gaps are occupied in the ground 
state configurations.}
\label{spectra}
\end{figure}

  The presence of the island of stability of SHN is due to large shell 
gaps in the single-particle spectra. The neutron and proton single-particle 
spectra obtained in spherical relativistic mean field (RMF) calculations of the $^{292}$120 and $^{304}$120 
nuclei are shown in Fig.\ \ref{spectra}\footnote{Similar figures are presented 
for some other CEDF's in  Refs.\ \cite{LG.14,BRRMG.99}. Ref.\ \cite{BRRMG.99} 
also provides detailed comparison with non-relativistic
Skyrme DFT results.}.  In order to create a more 
representative statistical ensemble, the calculations have been performed with 
10 CEDF's. Amongst those are the CEDF's NL3*, DD-ME2, DD-ME$\delta$ and DD-PC1, 
the global performance of which has been studied in Ref.\ \cite{AARR.14}. One 
can see that the $Z=120$ and $N=172$ shell gaps are especially pronounced in the 
$^{292}$120 nucleus. This a consequence of the presence of central
depression in density distribution generated by a predominant 
occupation of the high-$j$ orbitals above the $^{208}$Pb nucleus
\cite{AF.05-dep}. The increase of neutron number from $N=172$ up to $N=184$ 
is associated with the occupation of low-$j$ neutron orbitals which leads to 
a flatter density distribution in the $N=184$ system  \cite{AF.05-dep}. As 
a consequence, the $Z=120$ and $N=184$ shell gaps are reduced and 
$N=184$ gap is increased. As one can see in Fig.\ \ref{spectra}, 
these are rather general features which are independent of the 
CEDF.

\begin{figure}[ht]
\centering
\includegraphics[angle=0,width=10.0cm]{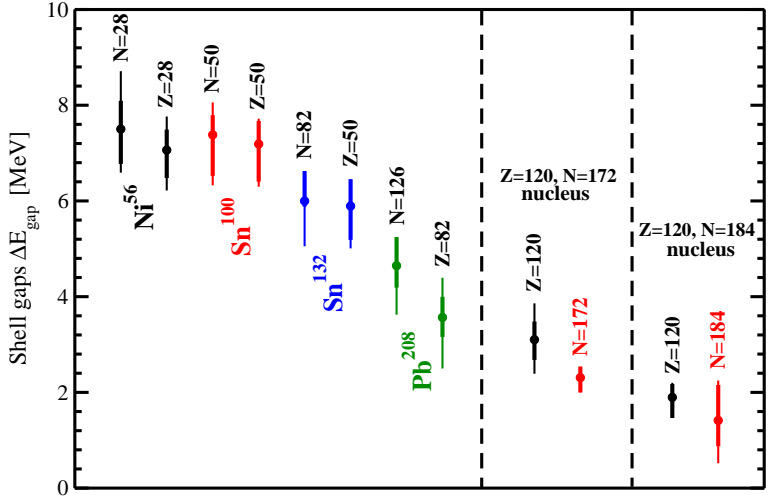}
\caption{Neutron and proton shell gaps $\Delta E_{\rm gap}$ of
the indicated nuclei. The average (among ten used CEDF's)
size of the shell gap is shown by a solid circle. Thin and thick
vertical lines are used to show the spread of the sizes of the
calculated shell gaps; the top and bottom of these lines
corresponds to the largest and smallest shell gaps amongst
the considered set of CEDF's. Thin lines show this spread for 
all employed CEDF's, while thick lines are used for the subset 
of four CEDF's (NL3*, DD-ME2,  DD-ME$\delta$ and DD-PC1). 
Particle numbers corresponding to the shell gaps are indicated.}
\label{gap-sizes}
\end{figure}

  Fig.\ \ref{spectra} clearly shows that there are theoretical 
uncertainties in the description of the energies of the single-particle 
states, their relative positions and the size of large shell gaps. The 
later is summarized in Fig.\ \ref{gap-sizes}, which shows the average 
sizes of these gaps and the spreads in their predictions. In addition, 
these gaps in SHN are also compared with the calculated gaps in the nuclei 
$^{56}$Ni, $^{100}$Sn, $^{132}$Sn and $^{208}$Pb. The general trend of the 
decrease of the size of the shell gaps with proton and neutron numbers 
are clearly visible. Definitely, the impact of theoretical uncertainties 
(shown by the spread of the sizes of the calculated gaps in Fig.\ 
\ref{gap-sizes}) on model predictions depends on relative sizes of 
theoretical uncertainties and calculated shell gaps. The presence of
theoretical uncertainties has less severe consequences on the predictions
of magic nuclei in $A\leq 208$  nuclei than on similar predictions for
SHN.

 The present analysis strongly suggests that in order to make reliable
predictions for SHN one needs a high predictive power for the energies 
of the single-particle states. It is also clear from Figs.\ \ref{spectra}
and \ref{gap-sizes} that the improvement in the DFT description of the 
energies of the single-particle states in known nuclei will also reduce 
the uncertainties in the prediction of the shell structure of SHN.
Unfortunately, the detailed investigation of the single-particle degrees
of freedom in the CDFT framework is in initial stage. This is because the
coupling of the single-particle motion with vibrations has to be taken 
into account (especially in spherical nuclei). So far, the accuracy of 
the description of the energies of the single-particle states and the 
sizes of shell gap in spherical nuclei has been studied in relativistic 
particle-vibration \cite{LA.11}  and quasiparticle-vibration \cite{Afanasjev2014_arXiv1409.4855} 
coupling models with the CEDF NL3* \cite{NL3*} only. The 
experimentally known gaps of $^{56}$Ni, $^{132}$Sn and $^{208}$Pb are 
reasonably well described in the relativistic particle-vibration calculations 
of Ref.\ \cite{LA.11}. The impact of particle-vibration on spherical shell 
gaps in  SHN has been investigated in Refs.\ \cite{LA.11,L.12}. Although 
particle-vibration coupling decreases the size of shell gaps, the $Z=120$ 
gap still remains reasonably large but there is a competition between smaller 
$N=172$ and $N=184$ gaps. The accuracy of the description of the energies of 
one-quasiparticle  deformed states in the rare-earth region and actinides 
has been statistically evaluated in Ref.\ \cite{AS.11} within the framework
of relativistic Hartree-Bogoliubov theory. On the one hand, these studies
have proved a success of CDFT; the covariant functionals provide a reasonable
description of the single-particle properties despite the fact that such observables
were not used in their fit. On the other hand, they illustrate the need for a
better description of the single-particle energies.

%%%%%%%%%%%%%%%%%%%%%%%%%%%%%%%%%%%%%%%%%%%%%%%%%%%%%%%%%%%%%%
%%%%%%%%%%%%%%%%%%%%%%%%%%%%%%%%%%%%%%%%%%%%%%%%%%%%%%%%%%%%%%%%

%%%%%%%%%%%%%%%%%%%%%%%%%%%%%%%%%%%%%%%%%%%%%%%%%%%%%%%%%%%%%%
%%%%%%%%%%%%%%%%%%%%%%%%%%%%%%%%%%%%%%%%%%%%%%%%%%%%%%%%%%%%%

%%%%%%%%%%%%%%%%%%%%%%%%%%%%%%%%%%%%%%%%%%%%%%%%%%%%%%%%%%%%%%
\begin{figure}[ht]
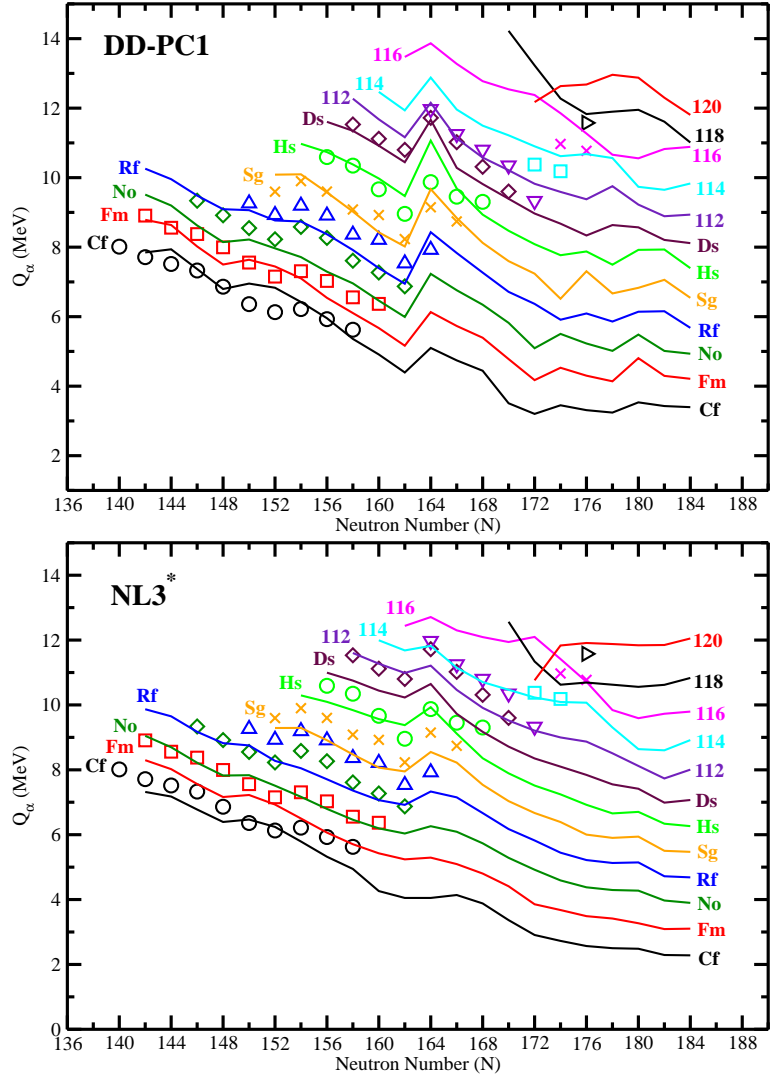

\centering
\includegraphics[angle=0,width=10.0cm]{fig-3-a.eps}
\includegraphics[angle=0,width=10.0cm]{fig-3-b.eps}
\caption{$Q_{\alpha}$ values of even-even superheavy elements as 
predicted by the RHB calculations with indicated CEDF's. The
formalism of Ref.\ \cite{AARR.14} is used in the RHB calculations. 
Experimental $Q_{\alpha}$ values are extracted from experimental
masses of Ref.\ \protect\cite{AME2012}.}
\label{Q_alpha}
\end{figure}
%%%%%%%%%%%%%%%%%%%%%%%%%%%%%%%%%%%%%%%%%%%%%%%%%%%%%%%%%%%%%

%%%%%%%%%%%%%%%%%%%%%%%%%%%%%%%%%%%%%%%%%%%%%%%%%%%%%%%%%%%%%%
\begin{figure}[ht]
\centering
\includegraphics[angle=-90,width=11.0cm]{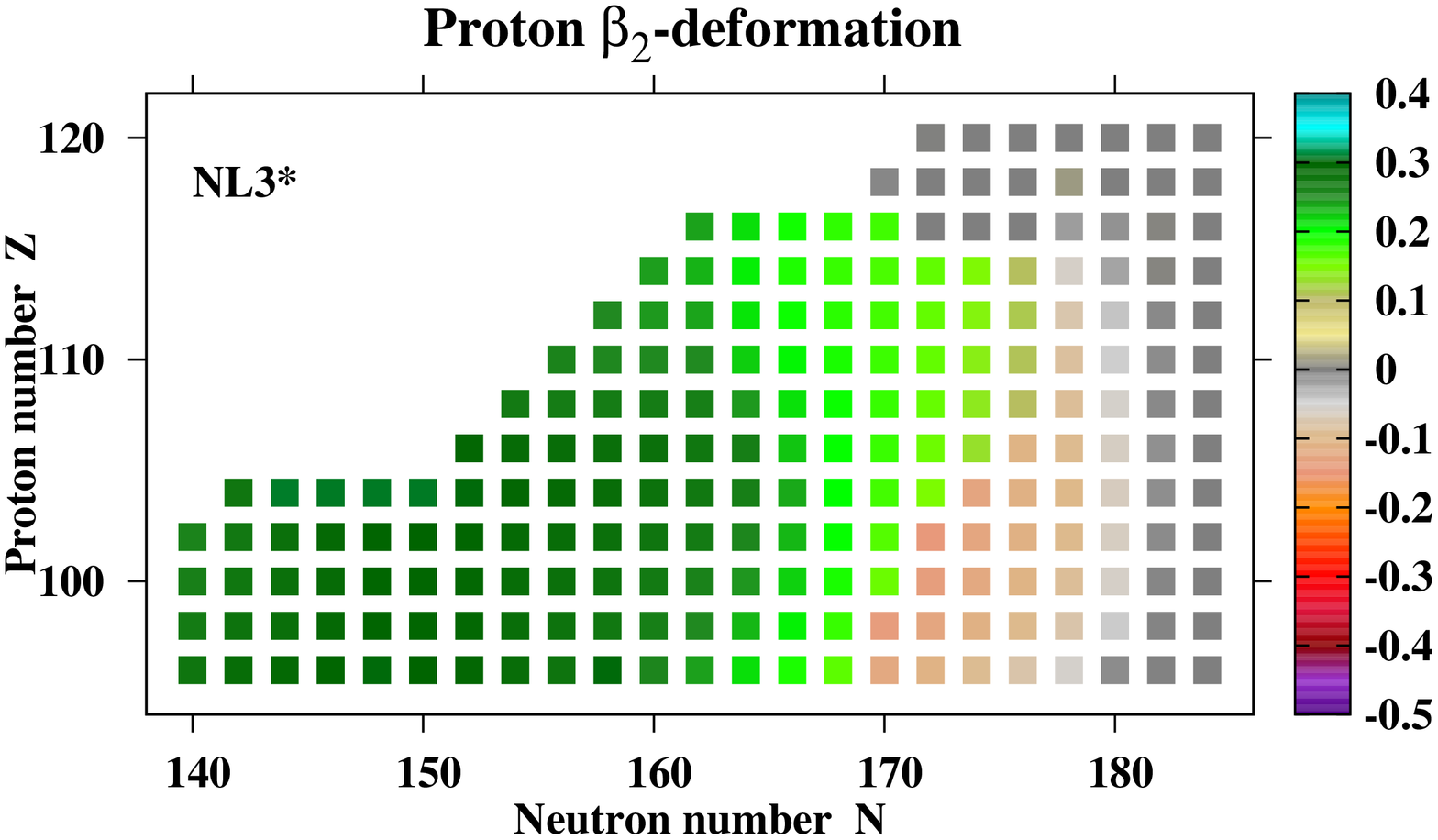}
\includegraphics[angle=-90,width=11.0cm]{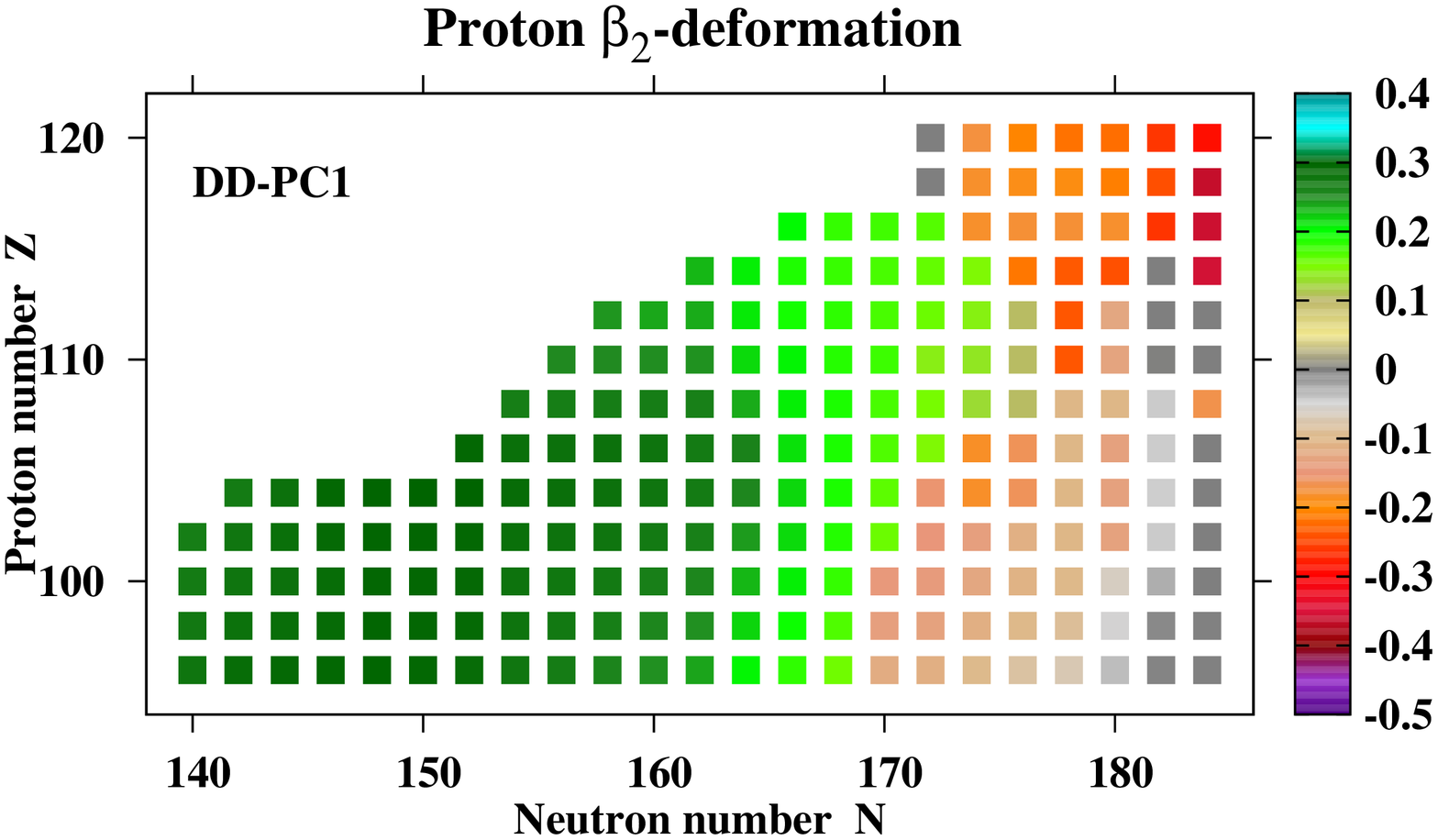}
\caption{Charge quadrupole deformations $\beta_2$ obtained in the 
         RHB calculations with indicated CEDF's.}
\label{deformation}
\end{figure}
%%%%%%%%%%%%%%%%%%%%%%%%%%%%%%%%%%%%%%%%%%%%%%%%%%%%%%%%%%%%

%%%%%%%%%%%%%%%%%%%%%%%%%%%%%%%%%%%%%%%%%%%%%%%%%%%%%%%%%%%%%%%%%%%%%%%%%%%%%%%%%%%%
\section{The $\alpha$-decay properties and the deformations of the ground states}
%%%%%%%%%%%%%%%%%%%%%%%%%%%%%%%%%%%%%%%%%%%%%%%%%%%%%%%%%%%%%%%%%%%%%%%%%%%%%%%%%%%%

  In superheavy nuclei spontaneous fission and $\alpha$ emission compete 
and shortest half-live determines the dominant decay channel  and the total 
half-live. Only in the case when spontaneous fission half-live of the nucleus 
is longer than half-live of $\alpha$ emission then superheavy nuclei can be 
observed in experiment. In addition, only nuclei with half-lives longer than 
$\tau =10\mu$s are observed in experiments.

  The $\alpha$ decay half-live depends on the $Q_{\alpha}$ values. The
$Q_{\alpha}$ values, obtained in the RHB calculations with the DD-PC1 
\cite{DD-PC1} and NL3* \cite{NL3*} CEDF's are compared with experimental
ones in Fig.\ \ref{Q_alpha}. One can see that reasonable agreement with
experimental data is achieved in both calculations. However, on average
somewhat better description is obtained with DD-PC1 CEDF. This is
a consequence of different fitting protocols\footnote{The DD-PC1 CEDF 
has been fitted to 64 deformed nuclei in the rare-earth region and actinides
and NL3* to only 12 spherical nuclei  (see Sect. II of Ref.\ \cite{AARR.14} 
for a detailed comparison of these two CEDF's).} and the fact that the binding 
energies are better described in DD-PC1 \cite{AARR.14}.

  The presence of the deformed $N=162$ shell gap reveals itself in the
presence of the peak at $N=164$ in the $Q_{\alpha}$ curves at fixed proton
number. The magnitude of this peak is dependent on the $N=162$ shell gap.
This peak is seen in experimental data of the Rf, Sg, Hs and Ds isotopic
chains. On average, the magnitude of this peak is somewhat underestimated
(overestimated) in the NL3* (DD-PC1) CEDF.

  The comparison of  experimental data with the calculated $Q_{\alpha}$ 
curves obtained in the CDFT (Fig.\ \ref{Q_alpha} in the present manuscript
and Fig. 18 in Ref.\ \cite{BHP.03}) and the ones obtained in non-relativistic 
models (see, for example, Fig. 18 in Ref.\ \cite{BHP.03} and Figs. 44 and 45 
in Ref.\ \cite{SP.07}) clearly indicate that available experimental data does not 
allow to distinquish the predictions of different models in respect of the 
position of the center of the island of stability. 

The calculated charge quadrupole deformations for these two CEDF's are 
plotted in Fig.\ \ref{deformation}. They reveal some interesting features 
which have not been discussed before. The $Z=120$ and $N=184$ SHN are 
spherical in the NL3* CEDF. On the contrary, the $Z=120, N\geq 174$ nuclei 
are oblate in the ground state in the DD-PC1 CEDF. This is in contradiction
with the expectations (based on large size of the $Z=120$ gap
(Fig.\ \ref{spectra})) that the $Z=120$ chain has to be spherical in the ground
states in both CEDF's.
%The analysis based on the 
%sizes of relevant gaps in the single-particle spectra (Fig.\ \ref{spectra}) 
%suggest that the isotopic $Z=120$ chain has to be spherical in both CEDF's
%based on the presence of large gap at $Z=120$. 
This result  clearly indicates 
that the softness of potential energy surface has to be taken into account 
when analyzing shell structure of SHN. Unfortunately, this fact is neglected 
in the analysis of shell structure of superheavy nuclei by means of so-called 
``two-nucleon shell gap''  in Refs.\ \cite{Rutz_PhysRevC.56.238.1997,LG.14} 
which is performed using the results of spherical calculations. 

%%%%%%%%%%%%%%%%%%%%%%%%%%%%%%%%%%%%%%%%%%%%%%%
\section{Fission barriers in superheavy nuclei}
%%%%%%%%%%%%%%%%%%%%%%%%%%%%%%%%%%%%%%%%%%%%%%%

  The properties of fission barriers is another important quantity 
which defines the stability of SHN. The systematic investigation of 
fission barriers in superheavy nuclei has been performed in 
the RMF+BCS framework with the NL3* CEDF in Ref.\ \cite{AAR.12}. 
The presence of a doubly-humped fission barrier structure in SHN 
is an example of the most striking difference between the relativistic 
and non-relativistic calculations; no outer fission barrier appears 
in absolute majority of non-relativistic calculations in the $Z\geq 
110$ SHN. The inclusion of triaxiality or octupole deformation in the 
RMF+BCS calculations always lowers (by around 2 MeV in the majority 
of the nuclei) the outer fission barrier as compared with the results 
of axially symmetric calculations. The underlying shell structure 
clearly defines which of the outer fission barrier saddle points 
(triaxial or octupole  deformed) is lower in energy. For example, the lowest saddle point is 
obtained in triaxial calculations in proton-rich nuclei with 
$N < 174$ (Ref.\ \cite{AAR.12}). On the contrary, the lowest saddle 
point is obtained in octupole deformed calculations in neutron-rich  
nuclei with $N > 174$.

\begin{figure}[ht]
\centering
\includegraphics[angle=0,width=12.6cm]{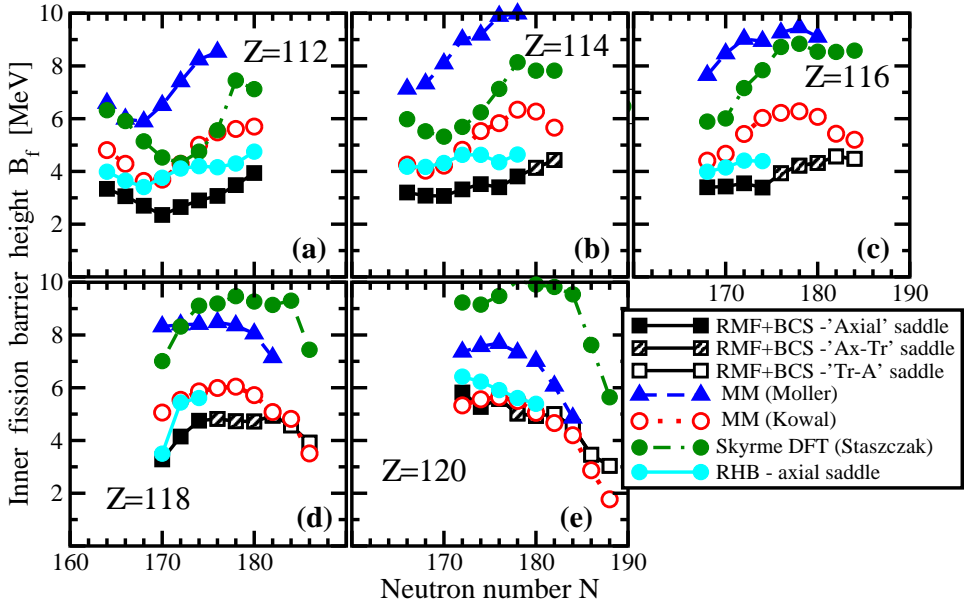}
\caption{Inner fission barrier heights $B_f$ as a function
of neutron number $N$. The results of the MM calculations 
are taken from Ref.\ \cite{MSI.09} (labeled as 'MM (M{\"o}ller)') 
and Ref.\ \cite{KJS.10} (labeled as 'MM (Kowal)'). The position 
of inner fission barrier saddle in deformation space varies as 
a function of particle number. The labeling of Ref.\ \cite{AAR.12} 
is used is order to indicate whether the saddle is axial (labeled as
'Axial'), has small  ($\gamma \sim 10^{\circ}$, labeled as 'Ax-Tr')
or large ($\gamma \sim 25^{\circ}$, labeled as 'Tr-A') $\gamma-$deformations 
in the RMF+BCS calculations. The results of Skyrme DFT calculations
with SkM* EDF have been taken from Ref.\ \cite{SBN.13}.}
\label{FB-SHE}
\end{figure}

  Fig.\ \ref{FB-SHE} shows how the models which have been benchmarked 
in a systematic way in the actinides extrapolate to the region of 
superheavy nuclei. These models describe inner fission barriers of 
actinides very accurately (see Fig.\ 2 in Ref.\ \cite{AAR.12}
and Ref.\ \cite{AAR.13-epj}). However, their predictions for SHN vary 
wildly; the difference in inner fission barrier heights between different 
models reaches 6 MeV in some nuclei. The more surprising fact is that 
the prediction of two macroscopic+microscopic (MM) models differ so 
substantially. As discussed in Ref.\ \cite{AAR.12} very limited 
experimental data on fission barriers in SHN is not reliable enough 
to distinquish between these model predictions.

 In addition to the RMF+BCS fission barriers we present also new results 
of axial RHB calculations (labeled as ``RHB - axial saddle''); the later 
are restricted to nuclei in which the saddle of inner fission barrier is 
axial in the RMF+BCS calculations. The principal difference between these two 
calculations lies in the treatment of pairing. The monopole pairing is
used in the RMF+BCS calculations \cite{AAR.12} and its strength is 
defined by the fit to ``empirical'' pairing gaps of Ref.\ \cite{MN.92}. On
the contrary, the separable pairing force of finite range is used in
the RHB calculations and its pairing strength is defined by the fit
to the moments of inertia in the actinides \cite{AO.13}. The differences
in calculated inner fission barriers are due to (i) different 
extrapolation properties of these two types of pairing on going from
actinides to superheavy region and (ii) the dependence of fission
barrier heights on the range (zero or finite) of pairing interaction 
\cite{KALR.10}. Because of these reasons the RHB results for the heights 
of inner fission barriers are higher than the RMF+BCS ones by 
roughly 1 MeV. Furthermore, they come closer to the 'MM (Kowal)' model 
predictions.

  Instead of fission barriers (which is indirectly measured quantity)
one can consider spontaneous fission half-lives $\tau_{SF}$ which is
directly measured quantity. However, the calculations of spontaneous 
fission half-lives represents a real challenge.  This is because there 
are significant uncertainties in $\tau_{SF}$ which emerge from different 
building blocks entering the standard semiclassical Wentzel-Kramers-Brillouin 
(WKB) formula \cite{BSDNS.11} which is used in the calculations of $\tau_{SF}$. 
These uncertainties have been analyzed in detail in Refs.\ \cite{SMBDNS.13,RR.14}.

  The calculated values of spontaneous fission half-lives also strongly depend 
to the underlying theory used to describe collective motion [typically the 
adiabatic time-dependent HFB (ATDHFB) or the generator coordinate method (GCM)]
and the approximations involved in the evaluation of the inertias; for a given
nucleus the difference between the $\tau_{SF}$ values calculated with ATDHFB and 
GCM can reach many orders of magnitude \cite{RR.14}. The $\tau_{SF}$ values also 
strongly depend on the poorly defined energy $E_0$ entering into action integral 
$S$; again the uncertainties reach several orders of magnitude.

  It was also shown in Ref.\ \cite{SMBDNS.13} that fission pathways strongly depend on 
assumptions underlying collective inertia. Perturbative cranking approximation, commonly 
used in ATDHFB, underestimates  the variations of mass parameters due to level crossings 
(configuration changes). As a result, a collective inertia drives dynamical fission path 
to near-axial shapes. When non-perturbative cranking inertia is employed, strong triaxiality 
is predicted for dynamical fission path in agreement with static calculations. So far this 
result has been obtained only for a single nucleus in  Ref.\ \cite{SMBDNS.13} and it remains 
to be seen whether it is a general conclusion.
  
    Unfortunately, no studies of spontaneous fission half-lives are available in the CDFT 
so far. This is contrary to the case of non-relativistic DFT's in which extensive studies  of 
spontaneous fission half-lives have been performed (see Refs.\ \cite{BSDNS.11,SMBDNS.13,RR.14} 
and reference therein). 

%%%%%%%%%%%%%%%%%%%%%%%%%%%%%%%%%%%%%%%%%%%%%%%%%%%%%%%%%%%%%%
%%%%%%%%%%%%%%%%%%%%%%%%%%%%%%%%%%%%%%%%%%%%%%%%%%%%%%%%%%%%%%%%%%%

%%%%%%%%%%%%%%%%%%%%%%%%%%%%%%%%%%%%%%%%%%%%%%%%%%%%%%%%%%%%%%%%%%%%%%%%%%
\section{Rotational excitations in actinides and light superheavy nuclei}
%%%%%%%%%%%%%%%%%%%%%%%%%%%%%%%%%%%%%%%%%%%%%%%%%%%%%%%%%%%%%%%%%%%%%%%%%%

%%%%%%%%%%%%%%%%%%%%%%%%%%%%%%%%%%%%%%%%%%%%%%%%%%%%%%%%%%%%%%%%%%%
\begin{figure}[ht]
\centering
\includegraphics[width=12.5cm,angle=0]{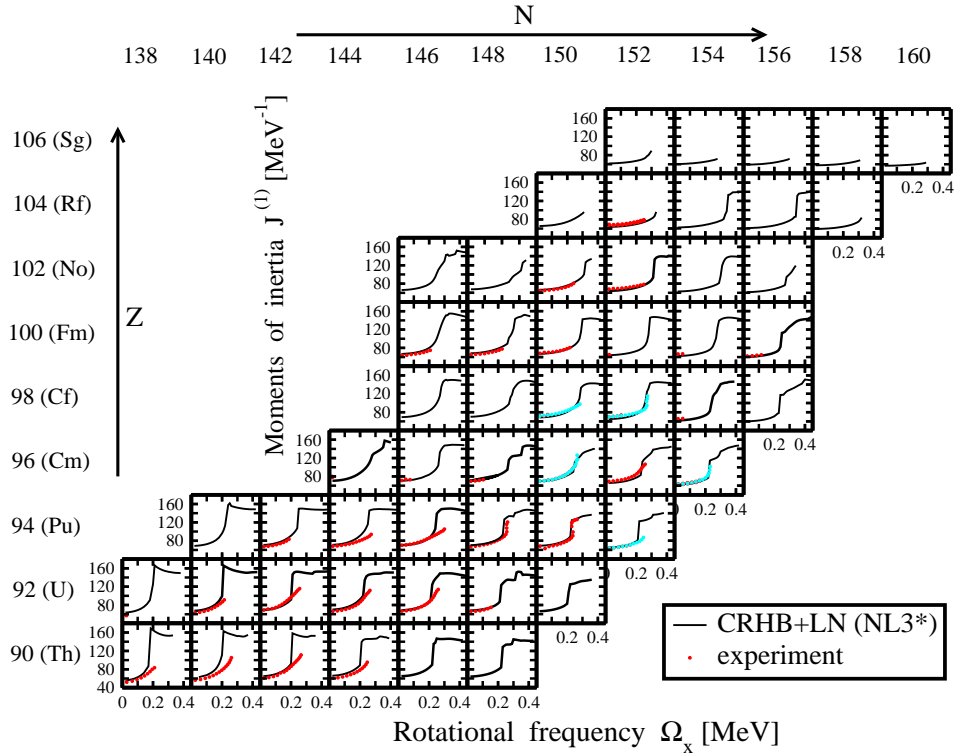}
\caption{ The experimental and calculated kinematic moments of inertia 
$J^{(1)}$ as a function of rotational frequency $\Omega_x$. The calculations 
are performed with the NL3* CEDF \cite{NL3*}. Calculated results and 
experimental data are shown by black lines and red dots, respectively. 
Cyan dots show new experimental data from Ref.\ \cite{Hota.thesis} 
which were not included in Ref.\ \cite{AO.13}. From Refs.\ 
\cite{AO.13,A.14}.} 
\label{sys-J1-NL3s}
\end{figure}
%%%%%%%%%%%%%%%%%%%%%%%%%%%%%%%%%%%%%%%%%%%%%%%%%%%%%%%%%%%%%%%%%%%

\begin{figure}[ht]
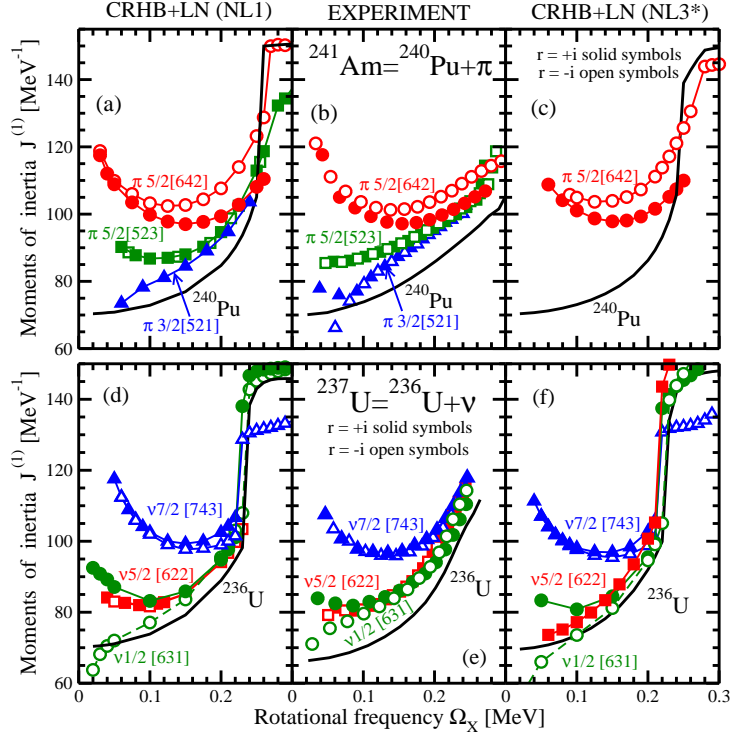

\centering
\includegraphics[width=9.5cm,angle=0]{fig-7-a.eps}
\includegraphics[width=9.5cm,angle=0]{fig-7-b.eps}
\caption{(top panels) Calculated and experimental kinematic moments
of inertia $J^{(1)}$ of the indicated one-quasiproton configurations
in the $^{241}$Am nucleus and ground state rotational band in
reference even-even $^{240}$Pu nucleus. Experimental data are shown in
the middle panel, while the results of the CRHB+LN calculations with the
NL1 and NL3* CEDF's in the left and right panels, respectively. 
The same symbols/lines are used for the same theoretical and experimental 
configurations. The symbols are used only for the configurations in odd-mass 
nucleus; the ground state rotational band in reference even-even nucleus 
is shown by solid black line. The label with the following structure
``Odd nucleus = reference even+even nucleus + proton($\pi$)/neutron($\nu$)''
is used in order to indicate the reference even-even nucleus and the type 
of the particle (proton or neutron) active in odd-mass nucleus. (bottom 
panels) The same as in top panels but for one-quasineutron configurations 
in $^{237}$U and ground state band in $^{236}$U. The experimental data are 
from Refs.\ \cite{Pu240,241Am-237Np}. Based on Ref.\ \cite{AO.13}.} 
\label{J1-241Am}
\end{figure}

  Figs.\ \ref{sys-J1-NL3s} and \ref{J1-241Am} show the results of the first 
ever (in any DFT framework) systematic investigation of rotational properties 
of even-even and odd-mass nuclei at normal deformation \cite{AO.13,A.14}. The 
calculations are performed within the CRHB+LN approach \cite{CRHB,A250}.
The gradual increases of the moments of inertia below band crossings are 
reproduced well. Either sharp or more gradual increases of the kinematic 
moments of inertia calculated at $\Omega_x \approx 0.2-0.30$ MeV are due 
to the alignments of the neutron $j_{15/2}$ and proton $i_{13/2}$ orbitals 
which in many cases take place at similar rotational frequencies. 
The upbendings observed in a number of rotational bands of even-even 
$A\geq 242$ nuclei are well described in model calculations (see Refs.\ 
\cite{AO.13,A.14} for details). However, the calculations also predict 
similar upbendings in lighter nuclei which have not been seen in experiment.
The stabilization  of octupole deformation at high spin, not included in the 
present CRHB+LN calculations, could be responsible for this discrepancy 
between theory and experiment \cite{AO.13}.

  The CRHB+LN approach provides much more consistent description 
of rotational properties in paired regime as compared with the cranked 
shell model plus particle-number conserving method (CSM+PNC) 
approach of Ref.\ \cite{ZHZZZ.12}. This is because it was necessary 
to adjust in the CSM+PNC approach the parameters of the Nilsson potential to
experimental single-particle energies, use experimental deformations
and employ different pairing gaps in even-even and odd-mass nuclei
in order to obtain comparable in accuracy with CRHB+LN approach the
description of experimental rotational properties of actinides 
\cite{A.14}.

%%%%%%%%%%%%%%%%%%%%%%%%%%%%%%%%%%%%%%%%%%%%%%%%%%%%%%%%%%%%%%%%%%%%%%%%
%%%%%%%%%%%%%%%%%%%%%%%%%%%%%%%%%%%%%%%%%%%%%%%%%%%%%%%%%%%%%%%%%%%%%%%%

  In the DFT framework, the description of rotational bands in odd-mass 
nuclei is more technically difficult than the one in  even-even nuclei. First, 
the effects of blocking due to odd particle have to be included in a fully 
self-consistent way which is done in the CRHB+LN computer code according 
to Refs.\ \cite{EMR.80,RS.80}. The blocking requires the identification of 
blocked orbital at all frequencies of interest and at all iterations which 
is non-trivial problem \cite{AO.13}. Second,  variational calculations with 
blocked orbital(s) are numerically less stable than the ones for the ground 
state bands in even-even nuclei because at each iteration of the variational 
procedure blocked orbital has to be properly identified. In general, the 
convergence depends on the interaction and relative energies of blocked 
orbital and its neighbour within a given parity/signature block (see Sect. V 
of Ref.\ \cite{AO.13}).

  A representative example of the CRHB+LN calculations for one-quasiparticle 
bands in $^{237}$Np and $^{241}$Am is shown in Fig.\ \ref{J1-241Am}; it comes 
from systematics of Ref.\ \cite{AO.13}. One can see that theoretical calculations 
describe well the absolute values of the kinematic moments of inertia of 
different one-quasiparticle configurations, their  evolution with rotational 
frequency, signature splitting and their relative properties with respect of 
the reference band in even-even nucleus. With few exceptions this is also 
true for other bands studied in Ref.\ \cite{AO.13}. Fig.\ \ref{J1-241Am} 
and Ref.\ \cite{AO.13}  also indicate that the results of the CRHB+LN calculations 
for a specific configuration only weakly depend on CEDF. The dependence of the 
convergence on the CEDF is clearly seen on the example of the $\pi 5/2[523]$ 
and $\pi 3/2[521]$ configurations in $^{241}$Am for which no convergence 
(convergence) has been obtained in the NL3* (NL1) CEDF.

  The systematic studies of Ref.\ \cite{AO.13} allowed to conclude that
rotational properties of one-quasiparticle configurations substantially
depend on the structure of blocked orbital. As a result, these properties 
reflected through the following fingerprints
\begin{itemize}
\item the presence or absence of signature splitting,

\item the relative properties of different configurations with 
respect of each other and/or with respect to the ground state
band in  reference  even-even nucleus,

\item the absolute values of the kinematic moments of inertia 
(especially at low rotational frequencies) and their evolution 
with rotational frequency

\end{itemize}
provide useful tools for quasiparticle configuration assignments.
Such configuration assignments are important, for example, for 
on-going experimental investigations of odd-mass light superheavy 
nuclei at the edge of the region where spectroscopic studies 
are still feasible (the nuclei with masses $A\sim 255$ and 
proton  number $Z\geq 102$) \cite{HG.08,AO.13}. However, it is 
necessary to recognize that the configuration assignment based 
on rotational properties has to be complemented by other independent 
methods and has to rely on sufficient experimental data \cite{AO.13}. 

%%%%%%%%%%%%%%%%%%%%%%%%%%%%%%%%%%%%%%%%%%%%%
\section{Conclusions}
%%%%%%%%%%%%%%%%%%%%%%%%%%%%%%%%%%%%%%%%%%%%%

  In conclusion, a short review of the recent progress in the study of actinides
and superheavy nuclei within covariant density functional theory has been 
presented.  It also includes new results displayed in Figs. 1-5 which
have not been published before. The uncertainties in the description of the 
energies of the single-particle states and the sizes of the shell gaps have 
been analyzed. Relatively small sizes of the shell gaps in the SHN imply that 
these  uncertainties can have a profound effect on the reliability of the
predictions. In such a situation other effects (such as softness of 
potential energy surface) have to be taken into account in analyzing
the shell structure of SHN. The differences in the predictions of the 
fission barriers of superheavy nuclei in different theoretical 
frameworks have been discussed. Finally, the accuracy of the description of 
rotational properties of actinides and superheavy nuclei and the possibility 
of their use for configuration assignment in odd-mass light superheavy nuclei 
have been analyzed.

%%%%%%%%%%%%%%%%%%%%%%%%%%%%%%%%%%%%%%%%%%%%%%%%%%%%%%%%%
\section{Acknowledgements}
%%%%%%%%%%%%%%%%%%%%%%%%%%%%%%%%%%%%%%%%%%%%%%%%%%%%%%%%%

 This work has been supported by the U.S. Department
of Energy under the grant DE-FG02-07ER41459.

\bibliographystyle{polonica}
\bibliography{refer-zak14}

\end{document}